\documentclass[12pt]{article}%
\usepackage{amsmath,latexsym}
\usepackage{graphicx}
\usepackage{amsmath}
\usepackage{amsfonts}
\usepackage{amssymb}%
\begin{document}

\title{{Charged wormholes in spacetimes of embedding
   class one}}
   \author{
Peter K. F. Kuhfittig*\\  \footnote{kuhfitti@msoe.edu}
 \small Department of Mathematics, Milwaukee School of
Engineering,\\
\small Milwaukee, Wisconsin 53202-3109, USA}

\date{}
 \maketitle

\begin{abstract}\noindent
The existence of charged black holes has
suggested that wormholes may also be charged.
The purpose of this paper is to construct a
general model of a charged wormhole that
proves to be a natural extension of the original
Morris-Thorne wormhole.  This goal is achieved
by means of the classical embedding theory
that has played a major role in the general
theory of relativity.   

\end{abstract}

\section{Introduction}\label{S:introduction}

Wormholes are handles or tunnels in spacetime
connecting widely separated regions of our
Universe or different universes in a
multiverse.  Apart from some forerunners,
macroscopic traversable wormholes were first
proposed by Morris and Thorne \cite{MT88}.
The wormhole geometry is described by the
following static and spherically symmetric
line element
\begin{equation}
  ds^{2}=-e^{\nu(r)}dt^{2}+\frac{dr^2}
  {1-\frac{b(r)}{r}}+r^{2}(d\theta^{2}
  +\text{sin}^{2}\theta\,d\phi^{2}),
\end{equation}
using units in which $c=G=1$.  Here $\nu=\nu(r)$
is called the \emph{redshift function}, which
must be everywhere finite to prevent the
appearance of an event horizon.  The function
$b=b(r)$ is commonly referred to the
\emph{shape function} since it determines the
spatial shape of the wormhole when viewed, for
example, in an embedding diagram \cite{MT88}.
The spherical surface $r=r_0$ is called the
\emph{throat} of the wormhole, where $b(r_0)
=r_0$.  The shape function must also meet the
requirements $b'(r_0)< 1$, called the
\emph{flare-out condition}., while $b(r)<r$
for $r>r_0$.  A final requirement is
asymptotic flatness: $\text{lim}_{r\rightarrow
\infty}\nu(r)=0$ and $\text{lim}_{r\rightarrow
\infty}b(r)/r=0$.

The flare-out condition can only be met by
violating the null energy condition (NEC)
\begin{equation}
  T_{\alpha\beta}k^{\alpha}k^{\beta}\ge 0
\end{equation}
for all null vectors $k^{\alpha}$, where
$T_{\alpha\beta}$ is the energy-momentum
tensor.  Matter that violates the NEC is
called ``exotic" in Ref. \cite{MT88} and
is usually confined to a narrow region
around the throat.  For the outgoing null
vector $(1,1,0,0)$, the violation of the
NEC becomes
\begin{equation}
   T_{\alpha\beta}k^{\alpha}k^{\beta}=
   \rho +p_r<0.
\end{equation}
Here $T^t_{\phantom{tt}t}=-\rho$ is the
energy density, $T^r_{\phantom{rr}r}= p_r$
is the radial pressure, and
$T^\theta_{\phantom{\theta\theta}\theta}=
T^\phi_{\phantom{\phi\phi}\phi}=p_t$ is
the lateral (transverse) pressure.

In the area of wormhole physics, an
interesting extension was proposed by Kim
and Lee \cite{KL01}.  Motivated by the
Reissner-Nordstr$\ddot{\text{o}}$m
spacetime, they considered the following
line element for a charged wormhole:
\begin{equation}
ds^{2}=-\left(1+\frac{Q^2}{r^2}
\right)dt^{2}+\frac{dr^2}
{1-\frac{b(r)}{r}+\frac{Q^2}{r^2}}
+r^{2}(d\theta^{2}
+\text{sin}^{2}\theta\,d\phi^{2}),
\end{equation}
where $Q$ is the electric charge.
The purpose of this paper is to
extend this special model to a general
Morris-Thorne wormhole with electric
charge by starting with a spacetime
of embedding class one, discussed in
the next section.  By assuming that
the redshift function is also dependent
on the charge $Q$, a possible line
element is
\begin{equation}\label{E:LINE}
ds^{2}=-e^{\nu(r,Q^2)}dt^{2}+\frac{dr^2}
{1-\frac{b(r)}{r}+\frac{Q^2}{r^2}}
+r^{2}(d\theta^{2}
+\text{sin}^{2}\theta\,d\phi^{2}).
\end{equation}
Even though the metric has changed, we
would like $b=b(r)$ to retain the usual
properties of a shape function in a
Morris-Thorne wormhole, although this
is not a requirement in Ref. \cite{KL01}.
We will see later, however, that due to
the embedding, the shape function
actually has the form $b=b(r,Q^2)$.
The line element will also be modified
to produce a model for a charged
wormhole that generalizes the original
Morris-Thorne wormhole in a natural way.

\section{The embedding}
Embedding theorems have played a major
role in the general theory of relativity,
as exemplified by the induced-matter
theory in Ref. \cite{pW15}: if the
embedding of spacetime is carried out
in accordance with Campbell's theorem,
then the resulting five-dimensional
theory can explain the origin of
matter.  Like any mathematical model,
the main criterion for its acceptance
is its usefulness.  In this paper, we
are primarily interested in spacetimes
of embedding class one, a special case
of the more general class $m$: an
$n$-dimensional Riemannian space is
said to be of embedding class $m$ if
$m+n$ is the lowest dimension of the
flat space in which the given space
can be embedded.  The resulting
mathematical model has proved to be
extremely useful in the study of compact
stellar objects
\cite{MG17, MM17, MRG17, sM17, sM16, sM19}
and will be used in this paper to extend
the above Kim-Lee model. To that end, we
first recall that the exterior
Schwarzschild solution is a Riemannian
space of embedding class two.  Following
Ref. \cite{MG17}, we will assume that
a spherically symmetric metric of
class two can be reduced to a metric of
class one by a suitable transformation of
coordinates.  To see how, we start with
the spherically symmetric line element
\begin{equation}\label{E:line1}
ds^{2}=e^{\nu(r)}dt^{2}-e^{\lambda(r)}dr^2
-r^{2}(d\theta^{2}+\text{sin}^{2}\theta\,
d\phi^{2}),
\end{equation}
where $\nu$ and $\lambda$ are
differentiable functions of the radial
coordinate $r$.  It is shown in Ref.
\cite{MG17} that this metric of class
two can be reduced to a metric of class
one and can therefore be embedded  in the
five-dimensional flat spacetime
\begin{equation}\label{E:line2}
   ds^2=-(dz^1)^2-(dz^2)^2-(dz^3)^2-
   (dz^4)^2+(dz^5)^2.
\end{equation}
This reduction can be accomplished by
the following transformation:
$z^1=r\,\text{sin}\,\theta\,
\text{cos}\,\phi$, $z^2=r\,\text{sin}\,\theta\,
\text{sin}\,\phi$, $z^3=r\,\text{cos}\,\theta$,
$z^4=\sqrt{K}\,e^{\nu/2}\,\text{cosh}
\frac{t}{\sqrt{K}}$, and $z^5=
\sqrt{K}\,e^{\nu/2}\,\text{sinh}
\frac{t}{\sqrt{K}}$.
The differentials of the components are:
\begin{equation}
  dz^1=\text{sin}\,\theta\,\text{cos}\,
    \phi\,dr+r\,\text{cos}\,\theta\,
    \text{cos}\,\phi\,d\theta-
    r\,\text{sin}\,\theta\,\text{sin}
    \,\phi\,d\phi,
\end{equation}
\begin{equation}
   dz^2=\text{sin}\,\theta\,\text{sin}
   \,\phi\,dr+r\,\text{cos}\,\theta\,
   \text{sin}\,\phi\,d\theta+r\,
   \text{sin}\,\theta\,\text{cos}
   \,\phi\,d\phi,
\end{equation}
\begin{equation}
  dz^3=\text{cos}\,\theta\,dr-r\,
  \text{sin}\,\theta\,d\theta,
\end{equation}
\begin{equation}
  dz^4=\sqrt{K}\,e^{\nu/2}\,\frac{\nu'}{2}
  \,\text{cosh}\,\frac{t}{\sqrt{K}}\,dr+
  e^{\nu/2}\,\text{sinh}\,\frac{t}
  {\sqrt{K}}\,dt,
\end{equation}
and
\begin{equation}
  dz^5=\sqrt{K}\,e^{\nu/2}\,\frac{\nu'}{2}
  \,\text{sinh}\,\frac{t}{\sqrt{K}}\,dr+
  e^{\nu/2}\,\text{cosh}\,\frac{t}
  {\sqrt{K}}\,dt.
\end{equation}
To facilitate the substitution into Eq.
(\ref{E:line2}), we first obtain the
expressions for $-(dz^1)^2-(dz^2)^2
-(dz^3)^2$ and $-(dz^4)^2+(dz^5)^2$:
\begin{equation}\label{E:part1}
  -(dz^1)^2-(dz^2)^2-(dz^3)^2=
  -dr^2-r^{2}(d\theta^{2}
  +\text{sin}^{2}\theta\,d\phi^{2})
\end{equation}
and
\begin{equation}\label{E:part2}
   -(dz^4)^2+(dz^5)^2=e^{\nu}dt^2-
   \frac{1}{4}Ke^{\nu}(\nu')^2\,dr^2.
\end{equation}
Substituting Eqs. (\ref{E:part1}) and
(\ref{E:part2}) in Eq. (\ref{E:line2}),
we obtain the new metric
\begin{equation}\label{E:line3}
ds^{2}=e^{\nu}dt^{2}
 -\left[1+\frac{1}{4}Ke^{\nu}(\nu')^2\right]dr^2
-r^{2}(d\theta^{2}+\text{sin}^{2}\theta\,
d\phi^{2}).
\end{equation}
So metric (\ref{E:line3}) is equivalent
to metric (\ref{E:line1}) if
\begin{equation}\label{E:key}
   e^{\lambda}=1+\frac{1}{4}Ke^{\nu}(\nu')^2,
\end{equation}
where $K>0$ is a free parameter.  Eq.
(\ref{E:key}) can also be obtained from
the Karmarkar condition \cite{kK48}
\begin{equation*}
  R_{1414}=
  \frac{R_{1212}R_{3434}+R_{1224}R_{1334}}
  {R_{2323}},\quad R_{2323}\neq 0,
\end{equation*}
which is equivalent to the above reduction.
In fact, Eq. (\ref{E:key}) is a solution
to the differential equation
\begin{equation*}
   \frac{\nu'\lambda'}{1-e^{\lambda}}=
   \nu'\lambda'-2\nu''-(\nu')^2,
\end{equation*}
readily solved by separation of
variables. So $K$ is actually an
integration constant \cite{MM17}.

Next, referring to line element
(\ref{E:line1}), to produce a wormhole
solution, we prefer the opposite
signature:
\begin{equation}
ds^{2}=-e^{\nu(r)}dt^{2}+e^{\lambda(r)}dr^2
+r^{2}(d\theta^{2}+\text{sin}^{2}\theta\,
d\phi^{2}),
\end{equation}
which takes us back to line element
(\ref{E:LINE}) with $b=b(r, Q^2)$:
\begin{equation}\label{E:line4}
ds^{2}=-e^{\nu(r,Q^2)}dt^{2}+\frac{dr^2}
{1-\frac{b(r,Q^2)}{r}+\frac{Q^2}{r^2}}
+r^{2}(d\theta^{2}
+\text{sin}^{2}\theta\,d\phi^{2}).
\end{equation}
As before, we assume that $\nu(r,Q^2)$
is a differentiable function of $r$ with
$\text{lim}_{r\rightarrow
\infty}\nu(r,Q^2)=0$.

\section{The wormhole solution}
From Eqs. (\ref{E:key}) and
(\ref{E:line4}), we obtain
\begin{equation}
   1-\frac{b(r,Q^2)}{r}+\frac{Q^2}{r^2}
   =\frac{1}{1+\frac{1}{4}Ke^{\nu(r,Q^2)}
   [\nu'(r,Q^2)]^2},
\end{equation}
where the prime denotes the derivative
with respect to $r$.  Solving for
$b(r,Q^2)$, we have
\begin{equation}\label{E:shape1}
   b(r,Q^2)=r\left(1+\frac{Q^2}{r^2}
   -\frac{1}{1+\frac{1}{4}Ke^{\nu(r,Q^2)}
   [\nu'(r,Q^2)]^2}\right).
\end{equation}
In line element (\ref{E:line4}), 
the effective shape function is given
by
\begin{equation}\label{E:eff}
   b_{\text{eff}}(r,Q^2)=b(r,Q^2)
   -\frac{Q^2}{r}.
\end{equation}
So, due the embedding,
\begin{equation}
   b_{\text{eff}}(r,Q^2)=
   r\left(1
   -\frac{1}{1+\frac{1}{4}Ke^{\nu(r,Q^2)}
   [\nu'(r,Q^2)]^2}\right).
\end{equation}
Unfortunately, the condition
$b_{\text{eff}}(r_0,Q^2) =r_0$ is now
impossible to meet.  It quickly becomes
apparent, however, that the effective
shape function in Eq. (\ref{E:eff})
implies that Eq. (\ref{E:shape1}) needs
a slight adjustment:
\begin{equation}\label{E:shape2}
   b(r,Q^2)=r\left(1+\frac{Q^2}{r^2}
   -\frac{1}{1+\frac{1}{4}Ke^{\nu(r,Q^2)}
   [\nu'(r,Q^2)]^2}\right)+\frac{Q^2}{r}.
\end{equation}
As noted in Sec. \ref{S:introduction},
we would like $b(r,Q^2)$ to satisfy all
the properties of a Morris-Thorne
wormhole.  This goal can be readily
achieved thanks to the free parameter
$K$ in the embedding theory.  In
particular, to satisfy the condition
$b(r_0,Q^2)=r_0$, we let
\begin{equation}
   K=\frac{\frac{r_0^2}{2Q^2}-1}
   {\frac{1}{4}e^{\nu(r_0,Q^2)}
   [\nu'(r_0,Q^2)]^2},\quad
   Q\neq 0.
\end{equation}
Then
\begin{multline}\label{E:shape3}
   b(r,Q^2)=r\left[1+\frac{Q^2}{r^2}
   \right.\\\left.
   -\left(1+\frac{\frac{r_0^2}{2Q^2}-1}
   {\frac{1}{4}e^{\nu(r_0,Q^2)}
   [\nu'(r_0,Q^2)]^2}\frac{1}{4}
   e^{\nu(r,Q^2}[\nu'(r,Q^2)]^2
   \right)^{-1}\right]
   +\frac{Q^2}{r}.
\end{multline}
It follows at once that $b(r_0,Q^2)
=r_0$.  Using this shape function, we
obtain a Morris-Thorne wormhole with
a nonzero electric charge.

As noted earlier, asymptotic flatness
requires that $\text{lim}_{r\rightarrow
\infty}\nu(r,Q^2)=0$.  Since $\nu(r,Q^2)$
is a differentiable function of $r$,
we also have $\text{lim}_{r\rightarrow
\infty}\nu'(r,Q^2)=0$.  Eq.
(\ref{E:shape3}) then yields the other
condition, i.e., $\text{lim}_{r\rightarrow
\infty}b(r,Q^2)/r=0$.

To check the flare-out condition at or
near the throat, we assume that $r\approx
r_0$ in Eq. (\ref{E:shape3}).  Then
\begin{equation}
   b(r,Q^2)\approx r\left(1+\frac{Q^2}
   {r^2}-\frac{2Q^2}{r_0^2}\right)
   +\frac{Q^2}{r}
\end{equation}
and
\begin{equation}
   b'(r_0,Q^2)\approx 1-\frac{4Q^2}
   {r_0^2}<1,
\end{equation}
provided that $r_0>2|Q|$, $Q\neq 0$.

For the final condition, $b(r,Q^2)<r$
near $r=r_0$, we simply let
$r_1\gtrsim r_0$ and observe that
\begin{equation}
   0<\frac{b(r_1,Q^2)}{r_1}\approx
   1+\frac{Q^2}{r_1^2}-\frac
   {2Q^2}{r_0^2}+\frac{Q^2}{r_1^2}
   <1
\end{equation}
since $Q\neq 0$.

In summary, a general model for a
charged wormhole is given by
\begin{equation}
  ds^{2}=-e^{\nu(r,Q^2)}dt^{2}+\frac{dr^2}
  {1-\frac{b(r,Q^2)}{r}}+r^{2}(d\theta^{2}
  +\text{sin}^{2}\theta\,d\phi^{2}),
\end{equation}
where $b(r,Q^2)$ is the shape
function in Eq. (\ref{E:shape3}).
The result is a natural
generalization of a Morris-Thorne
wormhole.

\section{Conclusions}
The existence of charged black holes
has suggested that wormholes may also be
charged.  This paper begins with a
discussion of Morris-Thorne wormholes,
followed by a charged wormhole model
due to Kim and Lee \cite{KL01}.  To
extend this special model, we made use
of the classical embedding theory that
is normally viewed as a viable and
effective mathematical model.  More
precisely, we made use of the fact
that a spherically symmetric metric
of class two can be reduced to a metric
of class one by a suitable transformation
of coordinates.  So the Kim-Lee model is
 not only extended, the embedding theory
 yields the following natural
 generalization of a Morris-Thorne
wormhole with electric charge:
\begin{equation*}
  ds^{2}=-e^{\nu(r,Q^2)}dt^{2}+\frac{dr^2}
  {1-\frac{b(r,Q^2)}{r}}+r^{2}(d\theta^{2}
  +\text{sin}^{2}\theta\,d\phi^{2}),
\end{equation*}
where
\begin{multline*}
   b(r,Q^2)=r\left[1+\frac{Q^2}{r^2}
   \right.\\\left.
   -\left(1+\frac{\frac{r_0^2}{2Q^2}-1}
   {\frac{1}{4}e^{\nu(r_0,Q^2)}
   [\nu'(r_0,Q^2)]^2}\frac{1}{4}
   e^{\nu(r,Q^2}[\nu'(r,Q^2)]^2
   \right)^{-1}\right]
   +\frac{Q^2}{r}
\end{multline*}
and $r_0>2|Q|$, $Q\neq 0$.  Both
the redshift and shape functions have
the required properties of a
Morris-Thorne wormhole, while the
wormhole spacetime itself is
asymptotically flat.

\end{document}